\documentclass[12pt]{article}
\usepackage{amsfonts}
\begin{document}

\baselineskip=20pt

\newfont{\elevenmib}{cmmib10 scaled\magstep1}
\newcommand{\preprint}{
   \begin{flushleft}
     \elevenmib Yukawa\, Institute\, Kyoto\\
   \end{flushleft}\vspace{-1.3cm}
   \begin{flushright}\normalsize  \sf
     YITP-03-50\\
     {\tt quant-ph/0308040} \\ August 2003
   \end{flushright}}
\newcommand{\Title}[1]{{\baselineskip=26pt
   \begin{center} \Large \bf #1 \\ \ \\ \end{center}}}
\newcommand{\Author}{\begin{center}
   \large  I.~Loris${}^{a,b}$ and R.~Sasaki${}^a$ \end{center}}
\newcommand{\Address}{\begin{center}
     ${}^a$ Yukawa Institute for Theoretical Physics,\\
     Kyoto University, Kyoto 606-8502, Japan\\
${}^b$ Dienst Theoretische Natuurkunde,
     Vrije Universiteit Brussel,   \\ 
Pleinlaan 2, B-1050 Brussels,  Belgium
   \end{center}}
\newcommand{\Accepted}[1]{\begin{center}
   {\large \sf #1}\\ \vspace{1mm}{\small \sf Accepted for Publication}
   \end{center}}

\preprint
\thispagestyle{empty}
\bigskip\bigskip\bigskip

\Title{ Quantum vs Classical  Mechanics:\\
role of elementary excitations       }
\Author

\Address
\vspace{1cm}

\begin{abstract}
Simple theorems relating a quantum mechanical system to
the corresponding classical one at equilibrium and connecting the quantum
eigenvalues to the frequencies of normal modes oscillations are presented.
Corresponding to each quantum eigenfunction, a `{\em classical
eigenfunction\/}' is  associated. Those belonging to `elementary excitations'
play an important role.
\end{abstract}

\section{Introduction}
\label{intro}
\setcounter{equation}{0}
In this paper we will address a general and a most fundamental
issue in multi-particle quantum mechanics; 
the correspondence/contrast between  quantum and classical mechanics.
Usually such correspondence/contrast is discussed in the 
``{\em quasi-classical\/}" or ``{\em quasi-macroscopic\/}"
regime of quantum mechanics in which the expectation values are good representations
of the classical variables.
These are exemplified in   Ehrenfest's 
well-known theorem or in the WKB 
method.
Here we ask a rather different question.
Suppose a multi-particle quantum mechanical system has a
unique ground state and {\em discrete\/} energy spectrum.
We naturally expect that the corresponding classical potential
has a well-defined minimum,
which gives an equilibrium point.
Near the equilibrium, the system is reduced to a collection of harmonic
oscillators as many as the degrees of freedom.
 
We will ask and answer in simple terms the following universal 
question in multi-particle quantum mechanics:
\begin{quote}
How can we relate the knowledge of the eigenfunctions and eigenvalues 
of a multi-particle quantum mechanical system to
the properties of the corresponding classical system, 
in particular, at equilibrium?
\end{quote}

In fact we show that for each quantum eigenfunction 
a corresponding ``{\em classical eigenfunction\/}" 
is defined as the $\hbar\to0$
limit, see (\ref{clasdef}).
These classical eigenfunctions 
satisfy simple eigenvalue equations (\ref{cleigen}).
Among them, there are ``{\em elementary excitations\/}", 
as many as the degrees of freedom,
corresponding to each {\em normal mode\/} of the small oscillations 
 at equilibrium (\ref{cleigen3}). 
They are the generators of the classical eigenfunctions.
The quantum eigenfunctions are the $\hbar$-deformations of 
these classical eigenfunctions.
We show that the main part of the generic quantum eigenvalues, 
which is proportional to Planck's constant $\hbar$,  
is given by a linear combination of the
(angular) frequencies of small oscillations 
with integer coefficients (\ref{linspec}).
 
Another motivation of the present research is to provide an analytical
proof for the recent results on the quantum vs classical integrability
in Calogero-Moser (C-M) systems by Corrigan-Sasaki \cite{cs}.
C-M systems \cite{Cal} are classical and quantum {\em integrable\/}
multi-particle dynamics based on root systems and the quantum eigenvalues
are expressed in terms of roots and weights. In other words, they have `integer'
energy spectra.
It was shown by direct numerical calculation \cite{cs} that most of the
classical data, for example, the (angular) frequencies of small oscillations
at equilibrium  are also `integers'. 
The Propositions 1--3 in section 2 give a simple analytic proof
for these interesting observations.
Thanks to the integrability (exact solvability), classical and quantum
eigenfunctions for C-M systems based on any root system can be constructed
explicitly.
They will be shown in a subsequent paper \cite{ls2} in some details, in particular, 
those for the elementary excitations. These  provide excellent explicit
examples of the main results of this paper.

The Planck's constant $\hbar$ is always written explicitly in this article.
This paper is organised as follows. In section 2, the formulation of multi-particle
quantum mechanics in terms of the {\em prepotential\/} is introduced and the 
basic results on the quantum and classical eigenfunctions are derived 
in an elementary way. 
Section 3 is devoted for summary and comments.

\section{Multi-particle Quantum Mechanics}
\label{qmechanics}

We will discuss a multi-particle quantum mechanical
system and its relationship with the corresponding
classical ($\hbar\to0$) dynamics. 
The dynamical variables 
are the coordinates
\(\{q_{j}|\,j=1,\ldots,r\}\) and their canonically conjugate momenta
\(\{p_{j}|\,j=1,\ldots,r\}\), subject to the Heisenberg commutation relations 
or the Poisson
bracket relations. 
We will adopt the standard vector notation in  \(\mathbb{R}^{r}\):
\[
   q=(q_{1},\ldots,q_{r}),\quad p=(p_{1},\ldots,p_{r}),
\quad q^2\equiv\sum_{j=1}^rq_j^2, 
\quad p^2\equiv\sum_{j=1}^rp_j^2, \ldots,
\]
in which $r$ is the number of particles.
In quantum theory, the momentum operator \(p_j\) acts as
a differential operator:
\[
   p_j=-i\hbar{\partial\over{\partial q_j}}, \quad j=1,\ldots,r.
\]

Throughout this paper we discuss the standard Hamiltonian system
\begin{equation}
H={1\over2}p^2+V(q),
\end{equation}
in which  we have assumed for simplicity that all the particles have the same mass,
which is rescaled to unity.
Let us start with  mild assumptions that the system has a 
unique and {\em square integrable\/} ground state $\psi_0$:
\begin{equation}
H\psi_0=0,\quad \int|\psi_0|^2\,d^r\!q<\infty,\quad E_0=0,
\end{equation}
and that it has  a finite (or an infinite) number of
{\em
discrete\/} eigenvalues:
\begin{equation}
H\psi_n=E_n\psi_n,\quad
E_n={\mathcal E}_n\hbar +\mathcal{O}(\hbar^2).
\label{eigexp}
\end{equation}
Here we adopt the convention that the ground state energy is vanishing $E_0=0$, by
adjusting the constant part of the potential $V$, see below.

Since the above time-independent Schr\"odinger equation is real
for a self-adjoint Hamiltonian and that the ground state has no {\em node\/} 
we express the ground state eigenfunction as
\begin{equation}
\psi_0(q)=e^{{1\over\hbar}W(q)},
\label{grstate}
\end{equation}
in which a real function $W=W(q)$ is called a {\em prepotential\/} \cite{bms}.
By simple differentiation of (\ref{grstate}), we obtain
\begin{equation}
p_j\psi_0=-i{\partial W\over{\partial q_j}}\psi_0,\quad
p^2\psi_0=-\sum_{j=1}^r\left[\left({\partial W\over{\partial q_j}}\right)^2
+\hbar{\partial^2 W\over{\partial q_j^2}}\right]\psi_0,
\end{equation}
which results in 
\begin{equation}
\left\{{1\over2}p^2+{1\over2}\sum_{j=1}^r\left[\left({\partial W\over{\partial
q_j}}\right)^2 +\hbar{\partial^2 W\over{\partial q_j^2}}\right]\right\}\psi_0=0.
\end{equation}
In other words, we can express the Hamiltonian and the potential  in terms of the prepotential
\cite{cs,bms}%
\footnote{
Similar formulas can be found within the context of supersymmetric quantum mechanics
\cite{khare}. Here we stress that supersymmetry is not necessary.}
\begin{equation}
H(W)={1\over2}p^2+V(q),\quad
V(q)={1\over2}\sum_{j=1}^r\left[\left({\partial W\over{\partial q_j}}\right)^2
+\hbar{\partial^2 W\over{\partial q_j^2}}\right].
\label{potform}
\end{equation}
By removing the obvious $\hbar$-dependent terms, let us define a {\em classical\/}
potential $V_C(q)$:
\begin{equation}
V_C(q)={1\over2}\sum_{j=1}^r\left({\partial W\over{\partial q_j}}
\right)^2.
\label{clpot}
\end{equation}
Equivalently one could introduce the classical Hamiltonian $H_C$ as an `average' of the
original Hamiltonian $H(W)$ with the one whose ground state is the
inverse of the original ground state%
\footnote{This is the main ingredient of the well-known Darboux transformation \cite{darb}.}
$H(-W)$:
\begin{equation}
H_C=\left(\vphantom{\mbox{\large I}}H(W)+H(-W)\right)/2={1\over2}p^2+V_C(q).
\label{clham}
\end{equation}
Conversely, (\ref{potform}) is a Riccati equation  determining the
prepotential $W$ for a given potential $V$ (or $V_C$).
Needless to say, it does not matter if the prepotential can be expressed in terms of
elementary functions or not.


\subsection{Equilibrium position and frequencies of small oscillations}

Now let us consider the equilibrium point of the  classical potential
$V_C$ (\ref{clpot}). The classical Hamiltonian (\ref{clham}) has a stationary
solution
at the classical equilibrium point, $p=0$,\ $q=\bar{q}$.
There could be, in general,  many stationary points of the classical potential $V_C$,
among which we will focus on the `{\em maximum}' point  $\bar{q}$
of the ground state
wavefunction $\psi_0$ \cite{cs}:
\begin{equation}
{\partial W(\bar{q})\over{\partial
q_j}}=0,\qquad \Longrightarrow
{\partial V_C(\bar{q})\over{\partial
q_j}}=\sum_{k=1}^r
{\partial^2 W(\bar{q})\over{\partial q_j\partial
q_k}}{\partial W(\bar{q})\over{\partial
q_k}}=0,\quad j=1,\ldots, r.
\label{Vmin}
\end{equation}
By expanding the classical potential $V_C$ around $\bar{q}$, we obtain
\begin{eqnarray}
V_C(q)&=&{1\over2}\sum_{j,\,k=1}^r
{\partial^2 V_C(\bar{q})\over{\partial q_j\partial
q_k}}(q-\bar{q})_j(q-\bar{q})_k +\mathcal{O}((q-\bar{q})^3)
\nonumber\\
&=&{1\over2}\sum_{j,\,k,\,l=1}^r{\partial^2 W(\bar{q})\over
{\partial q_j\partial q_l}}
{\partial^2 W(\bar{q})\over{\partial q_l\partial
q_k}}(q-\bar{q})_j(q-\bar{q})_k +\mathcal{O}((q-\bar{q})^3),
\end{eqnarray}
since $V_C(\bar{q})=0$, (\ref{clpot}).
Thus the eigen (angular) frequencies (frequency squared) of small oscillations
near the classical equilibrium are given as the eigenvalues of the Hessian matrix
$\widetilde{W}$ ($\widetilde{V}_C$):
\begin{equation}
\widetilde{W}=\mbox{Matrix}\left[\ 
{\partial^2 W(\bar{q})\over{\partial q_j\partial
q_k}}\right],\quad 
\widetilde{V}_C=\mbox{Matrix}\left[\ 
{\partial^2 V_C(\bar{q})\over{\partial q_j\partial
q_k}}\right]=\widetilde{W}^2.
\label{VWmat}
\end{equation}


\subsection{Quantum \& Classical Eigenfunctions}
\label{atsuteigfun}

Let us express the discrete eigenfunctions in  product forms
\begin{equation}
\psi_n(q)=\phi_n(q)\psi_0(q),\quad n=0,1,\ldots,\quad \phi_0\equiv1,
\end{equation}
in which $\phi_n$ obeys a simplified equation with the similarity transformed
Hamiltonian $\tilde{H}$ \cite{bms}:
\begin{eqnarray}
\tilde{H}\phi_n&=&E_n\phi_n,\label{sthameq}\\
\tilde{H}=e^{-{1\over\hbar}W}H e^{{1\over\hbar}W}
&=&-{\hbar^2\over2}\sum_{j=1}^r{\partial^2\over{\partial q_j^2}}-
\hbar\sum_{j=1}^r{\partial W\over{\partial q_j}}{\partial \over{\partial q_j}}.
\label{htilform}
\end{eqnarray}
Here we adjust the normalisation of  the eigenfunctions $\{\phi_n\}$
so that the corresponding ``{\em classical\/}"
eigenfunctions $\{\varphi_n\}$ are finite (non-vanishing) in the limit $\hbar\to0$:
\begin{equation}
\lim_{\hbar\to0}\phi_n(q)=\varphi_n(q),\quad n=1,2,\ldots, .
\label{clasdef}
\end{equation}
By taking the classical limit ($\hbar\to0$) of (\ref{sthameq}) and considering
(\ref{eigexp}), (\ref{htilform}), we arrive at an `{\em eigenvalue equation\/}' 
for the ``{\em classical\/}" wavefunctions
\begin{equation}
-\sum_{j=1}^r{\partial W\over{\partial q_j}}{\partial \varphi_n\over{\partial q_j}}
={\mathcal E}_n \varphi_n,\quad n=1,2,\ldots, .
\label{cleigen}
\end{equation}
Conversely one could define the classical eigenfunctions as  solutions of
the above eigenvalue equation.
In this case  the classical eigenfunctions must
satisfy
certain regularity  conditions. Then 
the {\em quantum\/} eigenfunction
$\phi_n$ could be considered as 
 an $\hbar$-{\em deformation\/} of the {\em classical\/}
eigenfunction
$\varphi_n$.
For the Calogero and Sutherland systems to be discussed in a subsequent paper \cite{ls2},
there is a one-to-one correspondence between the 
classical and quantum eigenfunctions.
For generic multi-particle quantum mechanical systems, the situation is less clear.

\bigskip

\subsection{Main Results}
\label{mainres}
The {\em classical\/}
eigenfunctions have the following remarkable properties:
\newtheorem{prop}{Proposition}
\begin{prop}
\label{prop21}
The product of two {\em classical\/}
eigenfunctions $(\varphi_n,{\mathcal E}_n)$ and $(\varphi_m,{\mathcal E}_m)$
is again a {\em classical\/}
eigenfunction with the eigenvalue ${\mathcal E}_n+{\mathcal E}_m$,
\begin{equation}
-\sum_{j=1}^r{\partial W\over{\partial q_j}}{\partial
(\varphi_n\varphi_m)\over{\partial q_j}} =({\mathcal E}_n+{\mathcal E}_m)
\varphi_n\varphi_m.
\label{cleigen2}
\end{equation}
\end{prop}
\begin{prop} 
\label{prop22}
The classical eigenfunctions vanish at the equilibrium 
$\bar{q}$
\begin{equation}
\varphi_n(\bar{q})=0, \quad n=1,2,\ldots, .
\end{equation}
\end{prop}
\begin{prop} 
\label{prop23}
The derivatives of a classical eigenfunction  at the equilibrium  
$\bar{q}$ form an eigenvector of the Hessian matrix $\widetilde{W}$,
iff $\left.\nabla \varphi_n\right|_{\bar{q}}\neq0$
\begin{eqnarray}
-\widetilde{W}\cdot\left.\nabla \varphi_n\right|_{\bar{q}}
&=&{\mathcal E}_n\left.\nabla \varphi_n\right|_{\bar{q}},
\quad n=1,2,\ldots, .
\label{cleigen3}\\
&\mbox{\rm or}&\nonumber\\
-\sum_{j=1}^r{\partial^2 W(\bar{q})\over{\partial q_k \partial
q_j}}{\partial
\varphi_n(\bar{q})\over{\partial q_j}} &=&{\mathcal E}_n
{\partial \varphi_n(\bar{q})\over{\partial q_k}},\quad
n=1,2,\ldots, .
\label{cleigen4}
\end{eqnarray}
\end{prop}

Obviously the Hessian matrix $\widetilde{W}$ (\ref{VWmat}) has at most $r$ different 
eigenvalues and eigenvectors.
The classical eigenfunctions $\{(\varphi_j,{\mathcal E}_j)\}$, $j=1,\ldots,r$ for
which
$\left.\nabla \varphi_j\right|_{\bar{q}}\neq0$ will be called ``{\em elementary
excitations\/}". At equilibrium, each corresponds to the {\em normal coordinate\/}
of the small oscillations with the eigen (angular) frequency ${\mathcal E}_j$.
That is, the `{\em main part\/}' $\mathcal{E}_n$ ({\em i.e,\/} $\mathcal{O}(\hbar)$ part)
(\ref{eigexp}) of the
quantum energy eigenvalue $E_n$ is given by the classic eigenfrequencies of the normal mode
oscillations at the classical equilibrium.  The
elementary excitations are the generators of the classical eigenfunctions. 
In other words, any classical eigenfunction can be expressed as
\begin{equation}
\varphi_1^{n_1}\cdots\varphi_r^{n_r},\quad {\mathcal E}=n_1{\mathcal E}_1+\cdots+
n_r{\mathcal E}_r,\quad n_j\in{\mathbb Z}_+,
\label{linspec}
\end{equation}
or a linear combination thereof with the same eigenvalue ${\mathcal E}$.
The above type of classical eigenfunctions  are obviously non-elementary and they 
have zero gradient at equilibrium, for example,
$\left.\nabla(\varphi_j\varphi_k)\right|_{\bar{q}}=0$.

\section{Summary and comments}

We have shown that for any multi-particle quantum mechanical system, 
\begin{quote}
the main part 
{\em i.e.\/} the $\mathcal{O}(\hbar)$ part, of the {\em quantum\/} energy eigenvalue is
determined solely by the corresponding {\em classical data\/}, {\em i.e.\/}
the eigenfrequencies of the normal mode oscillations at the classical equilibrium.
\end{quote}
This is a very powerful result, since for most multi-particle  systems
the quantum eigenvalues are hard to evaluate, whereas the eigenfrequencies
of the normal mode oscillations at classical equilibrium are easily
calculated. The Calogero-Moser (C-M) systems based on any root system \cite{Cal}
provide ideal explicit examples in which the above Propositions 1--3 are
thoroughly verified in Corrigan-Sasaki  paper
\cite{cs}. 
Thanks to the exact solvability, all the quantum eigenvalues of the C-M systems are 
known \cite{bms,HeOp} and they are compared with the eigenfrequencies
of the normal mode oscillations at classical equilibrium evaluated in 
\cite{cs}. The  classical and quantum eigenfunctions for the elementary excitations
are reported in some
detail in Loris-Sasaki  paper \cite{ls2}. It should be mentioned that Perelomov's recent work
\cite{perenew} asserts essentially our
Proposition 3 for the special cases of the
quantum-classical eigenvalue correspondence of the Sutherland systems. 

Let us present a few elementary examples of one degree of freedom quantum mechanics
to illustrate the prepotential method and the main results explicitly.

\underline{\bf Harmonic oscillator}\quad
The ground state wavefunction, the prepotential, the quantum and classical potential, 
etc are ($\omega>0$):
\begin{eqnarray}
\psi_0=e^{-\omega q^2/2\hbar},\quad W=-\omega q^2/2,\quad V(q)=\omega^2q^2/2-\omega\hbar/2,
\quad  V_C(q)=\omega^2q^2/2.
\end{eqnarray}
The quantum eigenvalues and eigenfunctions are:
\begin{eqnarray}
E_n=n\hbar\omega,\quad \mathcal{E}_n=n\omega, \quad \phi_n(q)=H_n(\sqrt{\omega/\hbar}\,q),
\end{eqnarray}
in which $H_n$ is the Hermite polynomial. The classical equilibrium point is the
origin $\bar{q}=0$ and the classical eigenfunction and the Hessian 
$\widetilde{W}$ (\ref{VWmat}) are: 
\begin{eqnarray}
\varphi_n(q)=\lim_{\hbar\to0}\hbar^{n/2}H_n(\sqrt{\omega/\hbar}\,q)=\omega^{n/2}q^n,\quad
-\widetilde{W}=\omega.
\end{eqnarray}
It is trivial to check  (\ref{cleigen}) and  the Propositions 1--3. \\
\underline{\bf `Soliton' potential}\quad
Let us consider a simple P\"oschl-Teller \cite{postell} potential
\begin{eqnarray}
\psi_0&=&1/(\cosh q)^{g/\hbar},\qquad  W=-g\log\cosh q,\nonumber\\ 
V(q)&=&-{g(g+\hbar)\over{2\cosh^2q}}+{g^2\over2},\quad
V_C(q)=-{g^2\over{2\cosh^2q}}+{g^2\over2}.
\label{solpot}
\end{eqnarray}
The quantum eigenvalues and eigenfunctions are:
\begin{eqnarray}
E_n=gn\hbar-n^2\hbar^2/2,\quad \mathcal{E}_n=gn,\
\phi_n(q)=(\cosh q)^n\,P_n^{(\alpha,\alpha)}(\tanh q),\ \alpha\equiv g/\hbar-n>0,
\end{eqnarray}
in which $P_n^{(\alpha,\beta)}(x)$ is the Jacobi polynomial of degree $n$.
The classical equilibrium point is the
origin $\bar{q}=0$ and the classical eigenfunction and the Hessian 
$\widetilde{W}$ (\ref{VWmat}) are: 
\begin{equation}
\varphi_n(q)=n!\lim_{\hbar\to0}\hbar^n\phi_n(q)=g^n(\sinh q)^n,\quad
-\widetilde{W}=g.
\end{equation}
It is easy to verify  (\ref{cleigen}) and the Propositions 1--3.
For integer  $g/\hbar$, $V(q)$ (\ref{solpot}) (without the constant term) is the {\em
reflectionless\/} potential corresponding to a KdV soliton. In both examples, $\varphi_1$
is the elementary excitation.

Throughout this Letter we have assumed that the prepotential $W$
is independent of the Planck's constant $\hbar$, for simplicity of the
presentation. The main content of this Letter is valid even if $W$ depends
on $\hbar$, so long as $\lim_{\hbar\to0}W=W_0$ is well-defined.
A celebrated example that $\lim_{\hbar\to0}W$ diverges is the hydrogen
atom, for which the classical equilibrium does not exist.
In this case the quantum-classical correspondence does not make sense
and the present formulation does not apply.

\section*{Acknowledgements}
We thank David Fairlie and Cosmas Zachos for useful discussion.
I.L. is a post-doctoral fellow with the F.W.O.-Vlaanderen (Belgium).


\end{document}